# Single wavelength 480 Gb/s direct detection over 80km SSMF enabled by Stokes Vector Kramers Kronig transceiver

Thang M. Hoang[1], Mohammed Y.S. Sowailem[1], Qunbi Zhuge[1,2,*], Zhenping Xing, Mohamed Morsy-Osman[1,3], Eslam El-Fiky[1], Sujie Fan[1], Meng Xiang[1], and David V. Plant[1]

[1]*Photonics System Group, Electrical and Computer Engineering Department, McGill University, 3480 University Street, Montreal, QC H3A 0E9, Canada*
[2] *Ciena Corporation, Ottawa, Ontario, K2K 0L1, Canada*
[3]*Elect. Eng. Dep., Alexandria University, Alexandria, Egypt*
*qunbi.zhuge@mcgill.ca*

**Abstract:** We propose 4D modulation with directed detection employing a novel Stokes-Vector Kramers-Kronig transceiver. It shows that employing Stokes vector receiver, transmitted digital carrier and Kramers-Kronig detection offers an effective way to de-rotate polarization multiplexed complex double side band signal without using a local oscillator at receiver. The impact of system parameters and configurations including carrier-to-signal-power ratio, guard band of the digital carrier, oversampling ratio and real MIMO is experimentally investigated. Finally, a record 480 Gb/s data rate over 80 km SSMF is achieved in a 60 Gbaud PDM-16QAM single carrier experiment with a BER below the threshold of $2.0 \times 10^{-2}$.

## 1. Introduction

Fiber optic communication traditionally drove the Internet through vast backbone networks spanned over hundreds and thousands of kilometers around the globe. Today, long-haul optical transmission is mainly based on high speed coherent technology, which can achieve 400 Gb/s and beyond per wavelength [1,2]. In the past five years, driven by the fast expansion of new emerging applications such as the Internet of Things (IoT), social network, virtual reality and cloud computing, data center (DC) interconnect traffic over short reach links is growing very rapidly. A 2016 Cisco report forecasts that in 2020 more than 80% of the global Internet traffic will be DC related, of which approximately 77% is predicted to stay within DC [3]. Such exponential growth of data center interconnect (DCI) traffic has driven the need for high-speed and cost-effective transceivers for 40~80 km short reach applications [4]. In these scenarios, direction detection solutions are currently more attractive than coherent solutions. Current generation switch in DC typically has 32 ports, each port utilizing an optical transceiver Quad small form-factor pluggable 28 (QSFP28). Each QPSFP28 transceiver provides a net rate of 4x25 Gb/s lanes. Higher bit rate employing pulse-amplitude modulation (PAM) multi-lane such as PAM4 8x50 Gb/s, is a potential solution for next generation 400 Gb/s transceiver. However, its spectral efficiency for C-band dense-wavelength-division-multiplexing (DWDM) transmission is low and optical chromatic dispersion (CD) compensation is needed [5- 8].

To further increase spectral efficiency, one potential solution is to utilize polarization-division-multiplexing (PDM) in direct detection based on a Stokes-vector receiver (SVR) [9-11]. Employing a SVR, two and three dimensional degrees of freedom (DOF) have been exploited and several high bit rate transmission demonstrations have been reported: 360-Gb/s at 20 km using PDM-2-ring/8PSK, 462-Gb/s at 1 km using PDM-PAM4+3PM, and 336-Gb/s at 80 km using PDM-16QAM-PAM4 [12-14]. Though the throughputs as well as number of DOFs/spectral efficiency are enhanced over the last three years, the use of SVR-based modulation is restricted by the three degrees of freedom in Stokes space [15-18].

Another active research topic is to employ single side band signaling in order to mitigate channel impairments and to take advantage of an extra DOF compared to conventional IM/DD [19, 20]. In 2016, a direct detection scheme based on Kramers Kronigs (KK) relation has been reported [21]. This approach enables the full complex field reconstruction after power detection by removing signal-to-signal beat interference (SSBI) with the assumption of minimum phase condition. KK has been shown with a superior performance over the reported SSBI cancellation techniques in the literature [21-24]. Using a KK detection, several high speed transmission have been demonstrated: 240-Gb/s/$\lambda$ signals over 125 km fiber using a heterodyne approach at receiver, 256-Gb/s/$\lambda$ signals over 160 km fiber aided with a digital carrier [23-24]. However, the spectral efficiency of these KK systems is limited by the single polarization operation or large guard band of carrier in heterodyne mode.

In this paper, we report the first single wavelength 400 Gb/s (raw 480 Gb/s) direct detection transmission over 80 km of standard single mode fiber (SSMF) using a novel Stokes Vector Kramers Kronig (SVKK) transceiver to achieve four-dimensional (4D) DOF modulation and digital chromatic dispersion (CD) compensation. The system is realized by sending a digital tone together with the PDM signals, and using the SVKK approach to de-multiplex polarizations in the Stokes space and then retrieve complex fields. Note that the digital carrier method is used in an independent work in [23]. Firstly, we demonstrate the feasibility of digital polarization de-multiplexing and study system performance at low baud rates (<30 GBaud) due to the lack of high speed analog-to-digital-converter (ADC) channels. It is shown that with different state-of-polarization rotations (SOP) at the receiver, the variation of received signal SNR is less than 0.2 dB. We also investigate the optimization of the carrier-to-signal-power-ratio (CSPR) and the guardband between the digital tone and the signal. In addition, other parameters are shown to have an impact on the KK performance including roll-off factor, sampling rate and CD precompensation. Lastly, we demonstrate dual-polarization signals at high baud rates (>30 GBaud). A 480 Gb/s single carrier signal with 60 Gbaud PDM-16QAM is successfully transmitted over 80 km SSMF with a bit-error-rate (BER) below the soft-decision forward error correction (SD-FEC) threshold of $2.0 \times 10^{-2}$ with a 20% overhead. Figure 1 shows the data rate and distance of this work relative to other reported direct detection schemes in the literature, so to the best of our knowledge this work reports the highest data rate per wavelength for >1 km direct detection transmissions.

The remainder of the manuscript is organized as follows. Section 2 presents the principle of the proposed SVKK scheme for 4D modulation direct detection. Section 3 details the experimental setup and the demonstration of SVKK based short reach transmission using DP-16QAM modulation. Section 4 concludes the paper.

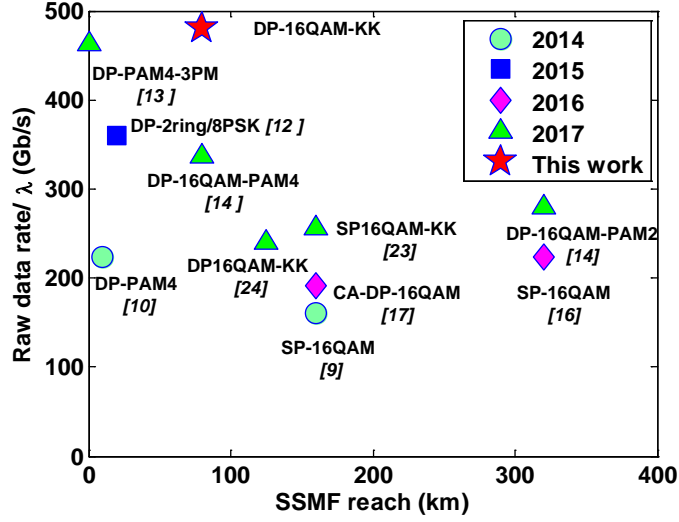

Fig. 1. Evolution of direct detection over the last four years.

## 2. Principle of 4D PDM SVKK transceiver

The architecture of the proposed SVKK transceiver is illustrated in Fig. 2 with digital signal processing (DSP) blocks. At the transmitter in Fig.2(a), complex double sideband QAM signals $E_X^t$ and $E_Y^t$ are generated after pulse shaping and pre-emphasis. Then a digital carrier is inserted at the right edge with a fixed guard band. It is noted that the use of digital carrier offers complexity reduction at transmitter side (no additional optical tone). A dual-polarization (DP) IQ modulator (IQM) is used to generate 4D optical signals. At the receiver, a SVR is employed to enable polarization de-multiplexing for PDM signals in the Stokes space [25, 26]. Denoting $S_2 = 2\text{Re}[E_{D+C}^{tX} E_{D+C}^{tY*}]$ and $S_3 = 2\text{Im}[E_{D+C}^{tX} E_{D+C}^{tY*}]$ as the real and imaginary parts of the beating between the two transmitted signals (data + carrier), respectively, and ignoring noise, the polarization rotation can be recovered as

$$\begin{bmatrix} |E_{D+C}^{tX}|^2 \\ |E_{D+C}^{tY}|^2 \\ S_2 \\ S_3 \end{bmatrix} = \begin{bmatrix} m_{11} & m_{12} & m_{13} & m_{14} \\ m_{21} & m_{22} & m_{23} & m_{24} \\ m_{31} & m_{32} & m_{33} & m_{34} \\ m_{41} & m_{42} & m_{43} & m_{44} \end{bmatrix} \begin{bmatrix} |E_X^r|^2 \\ |E_Y^r|^2 \\ 2\text{Re}[E_{D+C}^{rX} E_{D+C}^{rY*}] \\ 2\text{Im}[E_{D+C}^{rX} E_{D+C}^{rY*}] \end{bmatrix} \quad (1)$$

where $m_{11}, ..., m_{44}$ are 1-tap real values. $|E_X^r|^2$, $|E_Y^r|^2$, $S_2^r = \text{Re}[E_X^r E_Y^{r*}]$, and $S_3^r = \text{Im}[E_X^r E_Y^{r*}]$ are the four outputs of the SVR, the structure of which is plotted on the left side of Fig. 2(b). The de-rotation matrix in Eq. (1) can be obtained using either least mean square (LMS) adaptation or Stokes-orthogonal training samples [25, 26].

After the $|E_{D+C}^{tX}|^2$ and $|E_{D+C}^{tY}|^2$ are obtained, the KK detection is applied to these power signals to recover the complex field as [21]

$$\begin{bmatrix} E'^t_X \\ E'^t_Y \end{bmatrix} = \begin{bmatrix} |E_{D+C}^{tX}| e^{i\phi_X} - mean(|E_{D+C}^{tX}|^2) \\ |E_{D+C}^{tY}| e^{i\phi_Y} - mean(|E_{D+C}^{tY}|^2) \end{bmatrix} e^{-i\pi B t} \quad (2)$$

where $\phi = \frac{1}{2} \text{Im}\{\text{Hilbert}(\log(|E_{D+C}^t|^2))\}$, and B is the upconversion frequency determined by the digital carrier. The KK-detected $E'^t_X$ and $E'^t_Y$ in Eq. (2) are then low pass filtered to remove out of band noise.

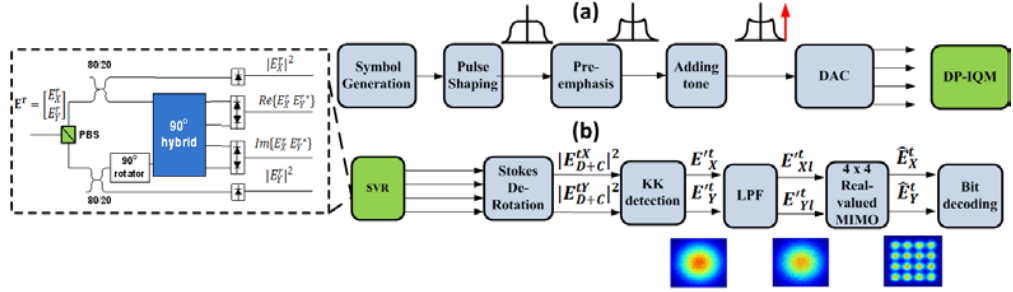

Fig. 2. Schematic diagram of the proposed SVKK system including (a) the transmitter with DSP blocks to generate signals, and (b) the receiver with DSP blocks after the SVR to recover PDM signals, and constellations of 60 Gbaud 16QAM signals after 80km.

The filtered $E'^t_{Xl}$ and $E'^t_{Yl}$ are fed into a 4x4 real MIMO LMS filter to remove linear intereference between $E^t_X$ and $E^t_Y$ and to mitigate the impact of imbalance (phase, amplitude and skew) among the four transmitted components (real and imaginary of X and Y polarizations) [27].

$$\begin{bmatrix} \text{Re}\{\hat{E}^t_X\} \\ \text{Im}\{\hat{E}^t_X\} \\ \text{Re}\{\hat{E}^t_Y\} \\ \text{Im}\{\hat{E}^t_Y\} \end{bmatrix} = \begin{bmatrix} h_{11} & h_{12} & h_{13} & h_{14} \\ h_{21} & h_{22} & h_{23} & h_{24} \\ h_{31} & h_{32} & h_{33} & h_{34} \\ h_{41} & h_{42} & h_{43} & h_{44} \end{bmatrix} \begin{bmatrix} \text{Re}\{E'^t_{Xl}\} \\ \text{Im}\{E'^t_{Xl}\} \\ \text{Re}\{E'^t_{Yl}\} \\ \text{Im}\{E'^t_{Yl}\} \end{bmatrix} \quad (3)$$

where $h_{11}, \ldots, h_{44}$ are the real taps. Note that in order to generate a digital carrier from the DACs without higher order harmonic and image components, the same pre-emphasis filter is applied to all four RF channels, which gives inherent imbalance at the transmitter side. In our system, CD compensation is also embedded into the MIMO filter to reduce complexity. Figure 2(b) shows the constellations of 60 Gbaud PDM-16QAM signals before (Gaussian-like shape due to CD) and after the MIMO filter, indicating its effectiveness. The outputs of Eq. (3) are the recovered data symbols, and the BER is counted based on the hard decision of those symbols.

## 3. Experimental setup and results

### 3.1 Experimental setup

The experimental setup is depicted in Fig. 3. The output of an external cavity laser operating at 1550.12 nm with 15.5 dBm optical power was fed into a DP-IQM with 35 GHz 3-dB bandwidth. Four 8-bit DACs operating at 88 GSps generated four electrical signals which were amplified using four discrete radio-frequency (RF) amplifiers. Each RF amplifier has 3-dB bandwidth of 50 GHz. The four outputs of RF amplifiers were used to drive the DP-IQM. The generated optical signal was passed through a 0.4 nm optical filter to remove the residual higher order harmonic components from the DACs. The output of this filter, of which the optical spectrum is plotted in Fig. 3, was boosted by an Erbium-doped fiber amplifier (EDFA) and the optical launch power was set to be 8 dBm using a variable optical attenuator (VOA) before the signal entered a spool of 80 km SSMF. After transmission, the signal was filtered by a 0.8 nm optical filter and amplified before the SVR, which is described in Fig. 2(b), with a received power of 12 dBm. The photodiodes (PDs) and balanced PDs had 3-dB bandwidths of 40 GHz. None of the PDs hosted trans-impedance amplifiers.

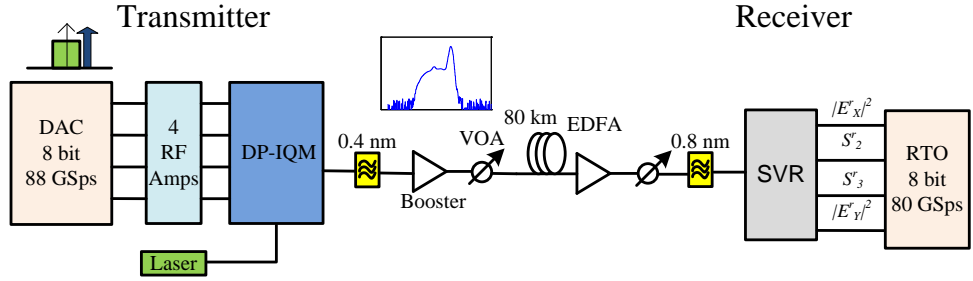

Fig. 3. Experimental setup. Inset is the optical spectrum of 60 Gbaud 16QAM after 0.4 nm optical filter.

The transceiver DSP blocks, which are described previously in Fig. 2, consisted of root-raised-cosine pulse shaping, digital carrier insertion with a predefined CSPR and guard band, and a MIMO filter as Eq. (3) with 61 real taps. The CSPR, guardband and roll-off factor were optimized to achieve the best performance.

### 3.2 Results

#### 3.2.1 System study and demonstration of polarization de-multiplexing with 27 Gbaud signals

In the first part of this experiment, we operated the SVKK system at low baud rates (<30 GBaud) to study the impact of CSPR and demonstrate the feasibility of polarization de-multiplexing. The received signals were digitized by the four 80 GS/s and 33 GHz ADC channels of one real-time oscilloscope (RTO). Figure 4(a) shows the impact of CSPR on 27 Gbaud 16QAM signals in a back-to-back (B2B) scenario. We can see that the optimal CSPR depends on the guard band, and generally it increases as the guard band (≥2 GHz) gets larger. This is because the DAC has limited bandwidth and finite frequency dependent effective number of bits (ENoB), which makes higher frequency digital tones noisier. For a 1 GHz guard band, there can be additional crosstalk between the signal and the tone. With a 11.5 dB CSPR and a 4 GHz guard band, we show in Fig. 4(b) the BER versus optical-signal-to-noise-ratio (OSNR) with 0.1 nm resolution at the B2B for 27 Gbaud 16QAM and 8QAM signals. It is noteworthy to mention that the proposed digital carrier approach has high transceiver noise (due to low electrical signal power) which can lead to an increased required OSNR in B2B.

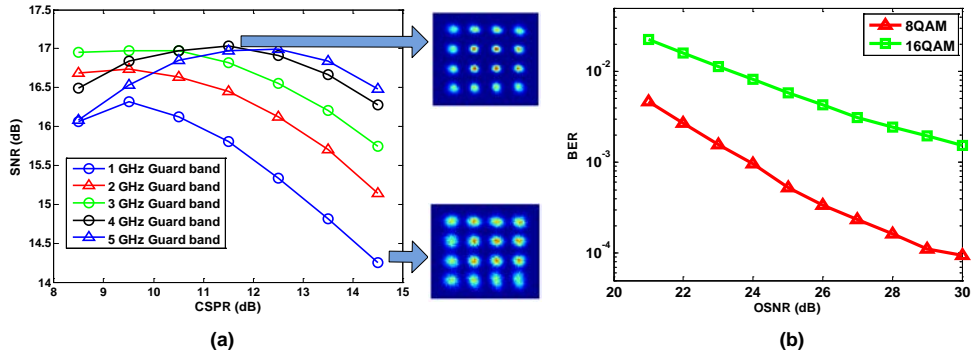

Fig. 4. (a) Impact of CSPR and Guardband with PDM-16QAM 27 Gbaud in B2B, (b) BER versus OSNR for 27Gbaud in B2B.

In order to perform polarization demultiplexing, we use training Stokes vector samples based on Eq. (1). During the training period, the digital carrier is not inserted. A periodic pattern of [1, 0, 1, 1] and [0, 1, 1, i] is applied for Ex and Ey, respectively, for estimating the polarization rotation matrix in Eq. (1). Similar approaches can be found in previously reported SVR systems [9, 26]. In Fig. 5(a), we show the synchronized training samples of transmitted $|E_X^t|^2$ and received $|E_X^r|^2$ in two incoming SOP scenarios at 80 km transmission. We can see that the training Stokes pattern is tolerant to the accumulated CD over 80 km SSMF. In this work, we use an averaging window of 128 samples centered in the middle of each time training slot to compute the de-rotation matrix in Eq. (1). In Fig. 5(b), we compare the performance of the proposed SVKK system with different states of polarization (SOP) of the 27 Gbaud 16QAM signal at the receiver. Results show that the signal-to-noise ratios (SNR) for the three SOPs, either at B2B or after 80 km, are less than 0.2 dB, which indicates the effectiveness of the SVKK polarization de-multiplexing.

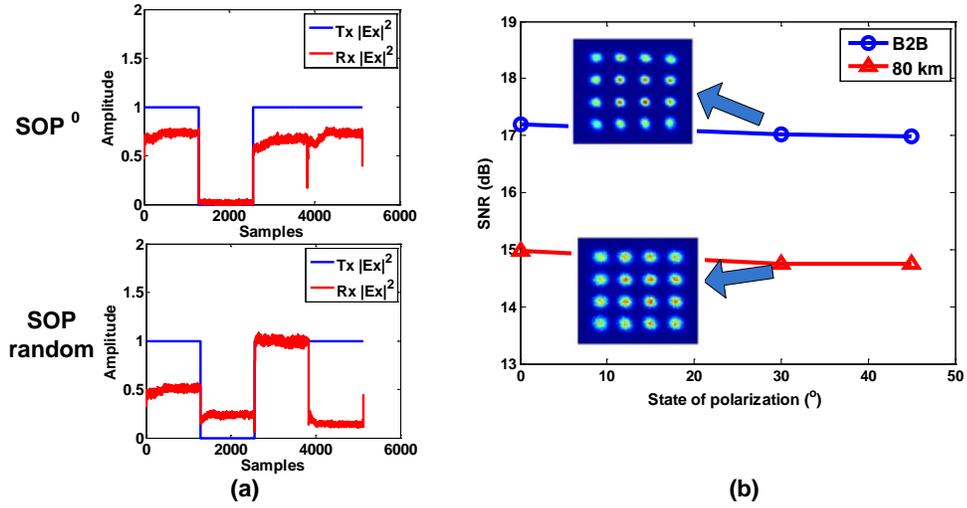

Fig. 5. Demonstration of Stokes polarization de-multiplexing: (a) the amplitude of transmitted $|Ex|^2$ and received $|Ex|^2$ during training period at 2 different SOP; (b) SNR versus SOP for 27 Gbaud DP-16QAM signals in B2B and 80 km cases.

In Fig. 6, we conduct a parametric study on the proposed SVKK approach with a CSPR of 11.5 dB. Figure 6(a) shows the BER versus baudrate using 8QAM, 16QAM, 32QAM and 64QAM in B2B. It indicates that 32QAM and 64QAM suffer from limited ENoB even at the low frequency regime. The effect is more severe in our approach since the digital carrier consumes significantly larger power compared to signal power. Figure 6(b) shows the impact of the roll-off factor (ROF) for 27 Gbaud 16QAM signals with a 4 GHz guardband in B2B. From this, the Nyquist-like pulse shaping, i.e. ROF close to 0, induces small penalty (~0.5 dB SNR). The explanation is that a Nyquist pulse has a higher peak-to-average-power ratio (PAPR) in time domain, which in turn may violate the minimum phase condition for the KK detection. Figure 6(c) shows the effectiveness of LMS in term of CD compensation at various oversampling ratio. It demonstrates similar performance between the three methods: LMS only, LMS + CD precompensation, and LMS+ CD postcompensation. It is noted that LMS (real MIMO) has to be used in the cases due to the inherent imbalance at the transmitter side. Figure 6(d) shows the recovered constellation using the conventional 2x2 butterfly complex MIMO and 4x4 real MIMO with a difference in SNR of ~ 1dB [28, 29].

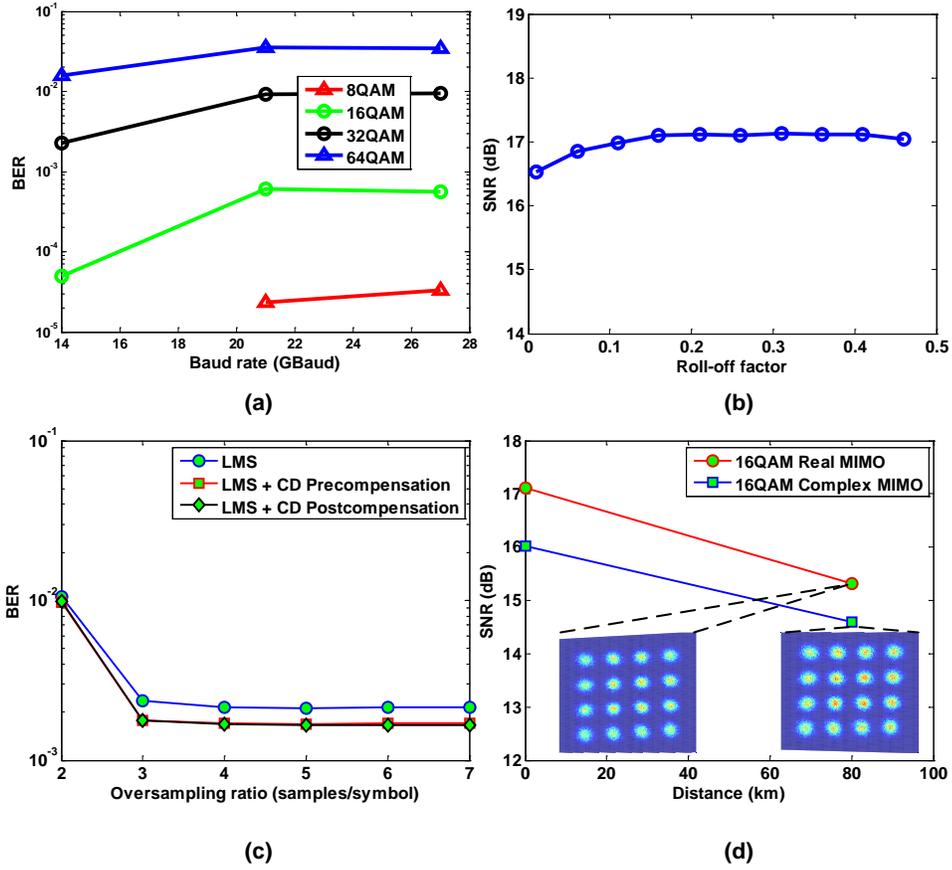

Fig. 6. Parametric study of SVKK (a) BER versus baudrate employing 8QAM, 16QAM, 32QAM and 64QAM in B2B, (b) SNR versus roll-off-factor for 16QAM 27 Gbaud in B2B, (c) BER versus oversampling ratio with three schemes for CD compensation, and (d) SNR versus distance in 16QAM 27 Gbau. Inset: 16QAM constellation after 80 km using real and complex MIMO.

### 3.2.2 *Demonstration of 480 Gb/s/λ transmission*

Since the feasibility of the polarization de-multiplexing, which is independent of baud rate, has been verified as shown in Fig. 5, in the second part of the experiment the SOP of the incident signal was aligned to avoid polarization mixing and the Stokes polarization de-multiplexing block in Fig. 2(b) was not employed. This was because we had to multiplex the four RTO channels to two channels with 160 GSps and 63 GHz bandwidth for the detection of the power terms $|E_X^r|^2$ and $|E_Y^r|^2$ at baud rates higher than 30 Gbaud. The CSPR and guard band were optimized for each baud rate. For 60 Gbaud, the CSPR was 13.5 dB and the guard band was 0.5 GHz. For 32.5 Gbaud, the CSPR was 12 dB and the guard band was 3.5 GHz. Figure 7(a) shows the BER versus baud rate for both the B2B and 80 km transmission scenarios. 32.5 Gbaud (raw bit rate 260 Gb/s) and 60 Gbaud (raw bit rate 480 Gb/s) PDM-16QAM signals were transmitted over 80 km with a BER below the hard-decision forward error correction (HD-FEC) threshold of $3.8\times10^{-3}$ and SD-FEC threshold of $2.0\times10^{-2}$, respectively. For the latter, excluding 20% FEC overhead, the net data rate is 400 Gb/s, which is the first single wavelength direct detection 400G system at 80 km, to the best of our knowledge. The oversampling ratio is an important parameter in the KK detection [21].

Figure 7(b) illustrates the impact of the oversampling ratio for signals with various baud rates. An important observation is that for any baud rate, the SNR is always suffered from a large penalty at 2 samples per symbol as already pointed out in [21]. This can be explained by Eq. (2), where a sufficient oversampling ratio( >3 samples/symbol) allows for an accurate waveform generation from nonlinear operations , i.e. square root and logarithm.

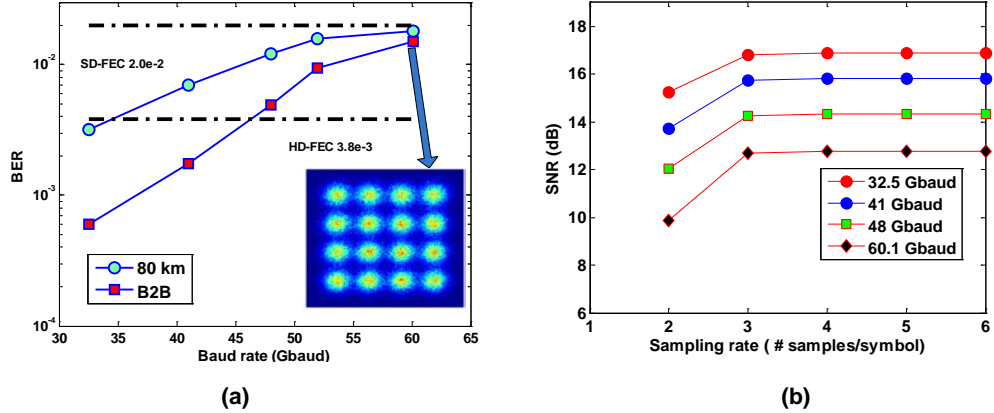

Fig. 7. (a) BER versus baudrate of DP-16QAM signals. Inset: constellation of 60 Gbaud signal after 80km, (b) SNR versus oversampling rate (samples/symbol) at various baudrate.

## 4. Conclusion

We have proposed a SVKK system using digital carrier to enable 4D modulation direct detection over dispersion-unmanaged C-band transmission. The feasibility of digital polarization de-multiplexing was experimentally demonstrated, and the impact of system parameters and configurations including carrier-to-signal-power ratio, guard band of the digital carrier, oversampling ratio and real MIMO was experimentally investigated. With this system, we reported the first single wavelength 480 Gb/s (net bit rate 400 Gb/s) transmission with 60 Gbaud PDM-16QAM signals over 80km SSMF in C-band.